# Accurate estimation of interfacial thermal conductance between silicon and diamond enabled by a machine learning interatomic potential

Ali Rajabpour[1,2,*], Bohayra Mortazavi[3], Pedram Mirchi[1], Julien El Hajj[2], Yangyu Guo[4], Xiaoying Zhuang[3], and Samy Merabia[2,*]

[1]Mechanical Engineering Department, Imam Khomeini International University, Qazvin 34148-96818, Iran
[2]Institut Lumière Matière, Université Claude Bernard Lyon 1-CNRS, Université de Lyon, Villeurbanne 69622, France
[3]Chair of Computational Science and Simulation Technology, Institute of Photonics, Department of Mathematics and Physics, Leibniz Universität Hannover, Appelstraße 11,30167 Hannover, Germany
[4]School of Energy Science and Engineering, Harbin Institute of Technology, Harbin 150001, China

## Abstract

Thermal management at silicon-diamond interface is critical for advancing high-performance electronic and optoelectronic devices. In this study, we calculate the interfacial thermal conductance between silicon and diamond using machine learning (ML) interatomic potentials trained on density functional theory (DFT) data. Using non-equilibrium molecular dynamics (NEMD) simulations, we compute the interfacial thermal conductance (ITC) for various system sizes. Our results show a closer agreement with experimental data than those obtained using traditional semi-empirical potentials such as Tersoff and Brenner which overestimate ITC by a factor of about 3. In addition, we analyze the frequency-dependent heat transfer spectrum, providing insights into the contributions of different phonon modes to the interfacial thermal conductance. The ML potential accurately captures the phonon dispersion relations and lifetimes, in good agreement with DFT calculations and experimental observations. It is shown that the Tersoff potential predicts higher phonon group velocities and phonon lifetimes compared to the DFT results. Furthermore, it predicts higher interfacial bonding strength, which is consistent with higher interfacial thermal conductance as compared to the ML potential. This study highlights the use of the ML interatomic potential to improve the accuracy and computational efficiency of thermal transport simulations in complex material systems.

## 1. Introduction

Diamond-silicon structures have attracted considerable attention as composite systems for applications in electronic and optoelectronic devices [1–5]. Chemical vapor deposition techniques provide a feasible method for synthesizing diamond films on silicon substrates. The increasing power densities and heterogeneity of modern electronic systems make interfacial thermal management critical to the performance and reliability of the main device. Efforts to minimize interfacial thermal resistance have considered various strategies, including nanostructuring of interfaces and enhancing interfacial bonding [6–8]. In the past decades, numerous studies have focused on the phononic properties of the diamond/silicon interface [9–13]. Using nanoscale graphoepitaxy to grow diamonds on silicon substrates shows tunable thermal transport across diamond membranes and diamond/silicon interfaces [11]. The effect of the thickness of the amorphous silicon interlayer on the thermal transport of silicon/diamond interfaces shows a significant increase in the interfacial thermal conductance with a thinner amorphous layer due to the reduction of the phonon mismatch and subsequent enhancement of the heat transport channels [14]. In addition, studies of the interfacial thermal conductance between diamond and silicon substrates have elucidated the temperature-dependent behavior and the effects of system size on thermal transport, as well as the significant influence of interface defects on phonon scattering and the consequent reduction in thermal conductance [12].

Molecular dynamics (MD) simulation is a powerful tool for studying interfacial thermal transport as it provides detailed insight into atomic-level interactions and transport mechanisms [15–20]. MD simulations can effectively model the complex dynamics of atoms and molecules at the interface with high resolution, capturing the interplay of phonons, defects, and interface structures that govern thermal transport [21]. By manipulating parameters such as interface roughness, interfacial bond strength, and temperature gradients, MD simulations can reveal how these factors influence thermal conductivity and interfacial thermal conductance (ITC). In addition, MD simulations allow the exploration of various scenarios that may be difficult or not practical to investigate experimentally. However, selecting an appropriate potential in MD simulations is critical as it directly affects the accuracy and reliability of the results. Different types of potentials, such as empirical potentials (e.g., Lennard-Jones), semi-empirical potentials (e.g., Tersoff), and *ab initio* potentials, offer different levels of accuracy and computational efficiency depending on the system being studied and the desired level of detail. Careful consideration must be given to selecting a potential that accurately represents the physical properties of materials while remaining computationally tractable for the intended simulation timescale and system size [22,23].

Machine learning potentials are increasingly recognized for their importance in MD simulations because they capture complex interactions with reasonable accuracy and efficiency [24–31]. Unlike empirical potentials, which rely on predefined functional forms and parameters, machine learning potentials are based on algorithms trained on large datasets of quantum mechanical calculations. This data-driven approach enables the representation of atomic interactions, including those involved in interfacial thermal transport. In other words, ML interatomic potentials offer significant advantages in terms of accuracy at the level of *ab initio* calculations but with computational times on the order of classical MD simulations.

Previous simulations of the silicon-diamond interface have mainly relied on semi-empirical potentials such as Tersoff and Brenner, which tend to overpredict the interfacial thermal conductance compared to experimental values [11,12,14]. In this study, non-equilibrium molecular dynamics simulation with machine-learning interatomic potentials are employed to accurately calculate the interfacial thermal conductance across bulk silicon and diamond structures. First, we detail the process of generating the potential by training it on density functional theory (DFT) data. We then perform simulations to calculate the interfacial heat transfer at various structure lengths, allowing us to evaluate the finite size effects on the thermal conductance. Finally, we analyze the results by presenting the frequency-dependent heat transfer spectrum, which provides insight into the thermal transport mechanisms at the silicon-diamond interface.

## 2. Methodology

In order to investigate heat transfer at the silicon/diamond interface, two silicon and diamond crystals with lattice constants, respectively 5.43 Å and 3.57 Å, are inserted and their interface is oriented along (1,0,0) direction. The initial distance between two crystals equals silicon and diamond's average atomic bond length. Also, in transverse directions, to minimize the stress at the interface, the value of 5.43 nm is considered for the system size along y and z dimensions. Figure 1 schematically shows the studied atomic structure in a limited length.

Tersoff interatomic potential is traditionally used to describe Si-Si, Si-C, and C-C covalent interactions in silicon, diamond, and their combinations. Instead of focusing on the semi-empirical Tersoff potential in this study, we intend to use potential training with quantum data using a machine-learning approach. For this purpose, the interface between silicon and diamond is first simulated by *ab initio* molecular dynamics simulation. The structure used for the *ab initio* simulation represent a total of 96 atoms.

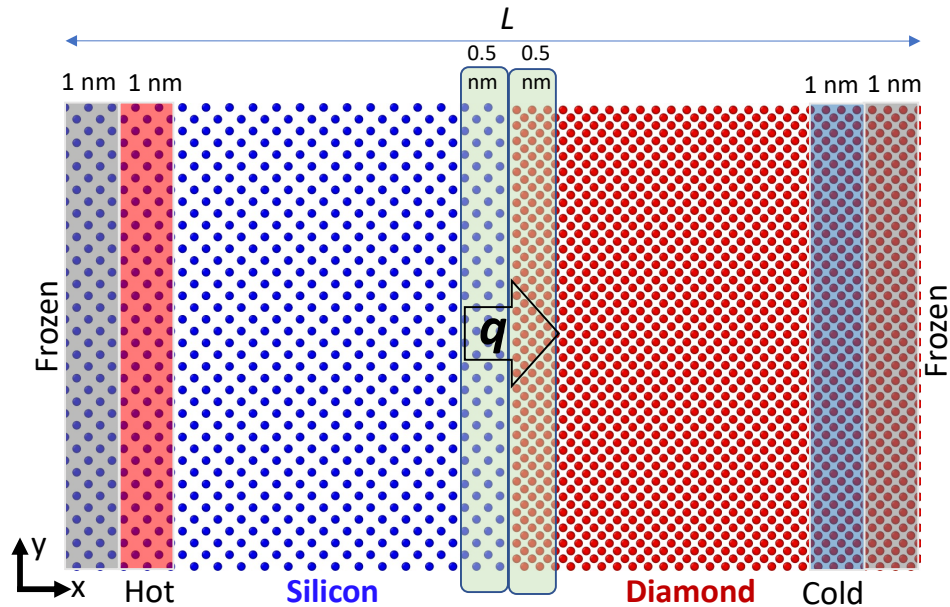

Fig1: Atomic structure of silicon/diamond interface and the NEMD setup to calculate the interfacial thermal conductance.

Vienna Ab-initio Simulation Package (VASP) [32] was utilized for ab-initio molecular dynamics (AIMD) simulations, employing the generalized gradient approximation (GGA) and revised Perdew-Burke-Ernzerhof (PBE) functional for solids [33]. The plane wave cutoff energy was set as 500 eV with a self-consistency convergence criterion of $10^{-4}$ eV, adopting a 2×2×2 **K**-point Monkhorst Pack mesh [34]. AIMD calculations were conducted for a periodic heterostructure with 96 atoms (40:Si and 56:C), pure diamond with 144 atoms, and pure silicon with 64 atoms, each for 3000 steps, with temperatures gradually increased from 1 to 2000 K. First the complete AIMD dataset with 9000 configurations was sampled, and 800 configurations were selected for training the first moment tensor potential (MTP) with 409 parameters and a cutoff distance of 4 Å, using the MLIP package [35]. Following the first training, the MTP accuracy was assessed across the entire AIMD dataset, identifying configurations with the highest extrapolation grades [36], which were added to the original dataset. The second and final MTP was trained from scratch using the updated dataset. This two-step passive fitting [37,38] procedure ensures the optimal usage of the training data for modeling thermal transport.

After obtaining the potential based on machine learning, we calculate the thermal conductance at the interface of the silicon/diamond structure using the non-equilibrium molecular dynamics (NEMD) simulation method. For this purpose, as shown in Figure 1, a known temperature gradient is applied to the system by creating hot and cold temperature baths at both ends of the system with target temperatures, and a heat flux is established across the structure. To create thermal insulating conditions at both ends of the system, atoms are frozen at the left and right boundaries. Periodic boundary conditions are applied along the y- and z-directions. The time step used throughout the simulation is 0.5 fs. The simulation runs for 50 ps to reach the relaxed state through the Nose-Hoover thermostat and the barostat. It then continues for 1 ns after applying the temperature gradient until the system reaches a steady state, and then continues for another 1 ns for data collection. All classical molecular dynamics simulations are performed using the LAMMPS package linked to the MLIP machine learning package [39,40].

Interfacial thermal conductance ($h$) is calculated based on the temperature difference on both sides of the interface through the equation $h = Q/(A.\Delta T)$ where $Q$ is the heat flow across the system, $A$ is the cross-sectional area, and $\Delta T$ is the temperature difference on the two sides of the interface. To calculate $\Delta T$, the extrapolation of the temperature profile on both sides of the interface is used discarding temperature data very close to the interface.

To calculate the frequency-dependent interfacial thermal conductance, the decomposition of the heat flux passing through the interface is considered:

$$Q = \int_0^\infty q(\omega)\frac{d\omega}{2\pi} = \sum_{i \in I}\sum_{j \in J}\int_0^\infty q_{i \to j}(\omega)\frac{d\omega}{2\pi} \tag{1}$$

where $\omega$ is the frequency, I and J represent the left and right sides of the interface (with a thickness equal to the interatomic force cutoff). The spectral heat flux can be calculated from the following equation [41–44]:

$$q(\omega) = 2 \, \text{Re} \, [ \, \sum_{j \in J} \int_0^\infty \langle \boldsymbol{F}_j(t) . \boldsymbol{v}_j(\boldsymbol{0}) \rangle \exp \, (-i\omega t) \, dt \, ] \qquad (2)$$

Here, Re is the real part, $\boldsymbol{F}_j = \sum_{i \in I} \boldsymbol{F}_{ij}(t)$ is the net force exerted on each atom at one side of the interface ($J$) only by the atoms on the other side of the interface. $\boldsymbol{v}$ is the atomic velocity, and $< \cdots >$ denotes an average over the simulation time. Having the spectral heat flux, the spectral conductance can be calculated from the following equation:

$$h(\omega) = \frac{q(\omega)}{A \, \Delta T} \qquad (3)$$

In classical molecular dynamics, the statistical distribution of kinetic energy obeys the equipartition theorem. However, at low temperatures (i.e., below the Debye temperature of the material), quantum effects become significant, and the Bose-Einstein distribution is more appropriate to describe the behavior of quasi-particles such as phonons. Therefore, a correction factor is introduced to adjust for the differences between the equipartition and Bose-Einstein distributions at each frequency to accurately model the system at any temperature. For this purpose, we apply a coefficient equal to the ratio of the quantum heat capacity ($\hbar\omega \, \partial f_{\text{BE}}/\partial T$) to the classical heat capacity ($k_B$) in a Landauer approach[44]:

$$h = \frac{1}{A \, \Delta T} \int_0^\infty q(\omega) \frac{\hbar\omega}{k_B} \frac{\partial f_{\text{BE}}}{\partial T} \frac{d\omega}{2\pi} \qquad (4)$$

where $f_{\text{BE}}$ is the Bose-Einstein distribution function and $\hbar$ is the reduced Planck constant.

## 3. Results & discussion

In this section, the results of the interfacial thermal conductance for different lengths of silicon/diamond heterostructures are presented, and the length-independent interfacial thermal conductance calculated by the ML potential is compared with the experimental results and also with other MD reports that consider Tersoff or Brenner potentials.

Figure 2(top panel) shows the temperature profile along the silicon/diamond heterostructure when the temperature of the hot thermostat is 320 K and the temperature of the cold thermostat is 280 K. The length of the considered structure is equal to 32 nm and the thickness of the hot/cold bath and the thermal insulating regions are 1 nm, individually. As can be seen in Fig. 2 (top panel), a significant temperature jump is observed at the interface. In addition, it is known that the temperature gradient on either side of the interface is proportional to the effective thermal conductivity of silicon and diamond, so the temperature gradient in silicon is higher due to its lower thermal conductivity than diamond. Using the extrapolation of the fitted lines to silicon and diamond temperature points, $\Delta T$ at the middle of the structure is calculated with a value equal to 28 K. To calculate the heat flux in the system; Figure 2 (bottom panel) is considered, which shows the accumulated variations of the energy of the hot and cold temperature thermostats with respect to the time in steady state. The slope of the energy change with respect to time in both hot and cold baths is equal to the heat flux established in the system. Also, the good matching of the rate of energy changes in the hot and cold baths (less than 5% difference) indicates the conservation of energy in the MD simulation with ML interatomic potential.

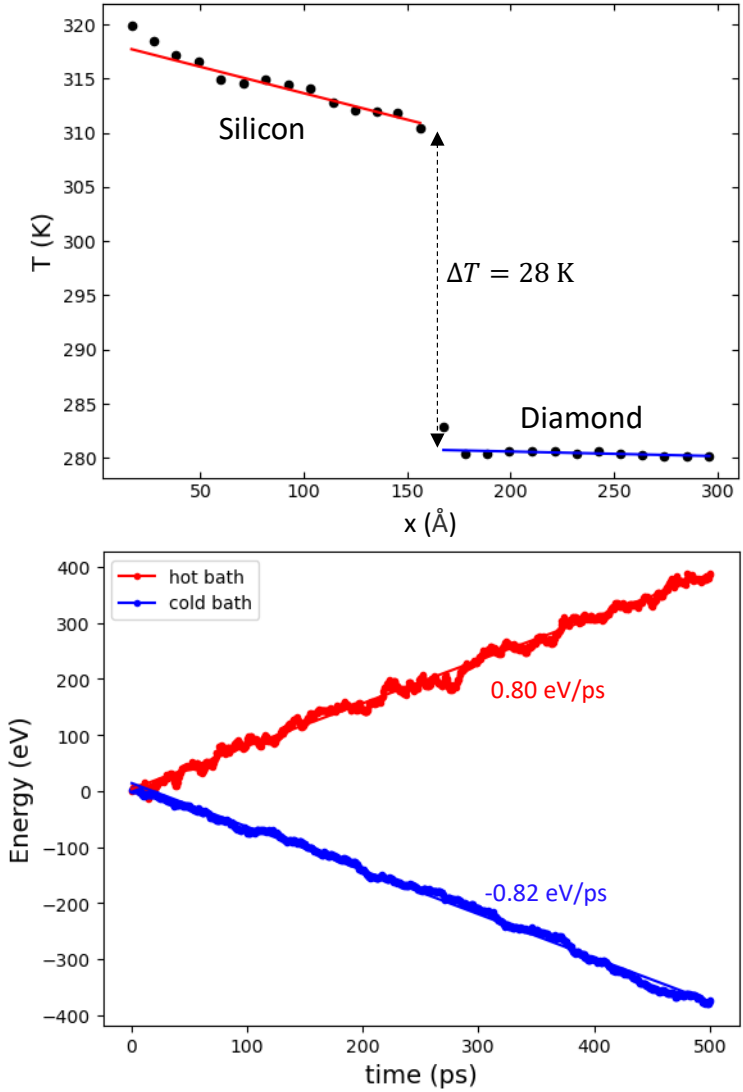

Fig2. **top panel:** Temperature profile at silicon/diamond heterostructure using ML potential; **bottom panel:** Accumulated energy variations of hot and cold thermostats with respect to the time

Figure 3(left panel) shows the interfacial thermal conductance for different heterostructure lengths. It can be seen that the interfacial thermal conductance for short heterostructure lengths is a function of the system size. There are several reasons for this size dependence. First, the thermostats may scatter the phonons traveling across the interfaces in small systems. Second, the contribution of ballistic phonons to thermal transport is more pronounced for system sizes smaller than the phonon mean free path. By considering a functional form $h(L) = aL/(b + L)$, the value of $h_\infty$ corresponding to the infinite length of the system can be estimated.

Figure 3(right panel) shows the results of $h_\infty$ calculated by the ML potential with and without quantum corrections. The ML potential results give a much closer estimate of the experimental results than the Tersoff potential. The relative difference between the ML potential and the experimental results is 10%, due to possible lattice defects in the experimental synthesis and approximations considered in the molecular dynamics simulations (such as the simple treatment of quantum corrections). In addition, it can be seen that Tersoff and Brenner potentials overestimate the interfacial thermal conductance by approximately 300% times higher than ML potential.

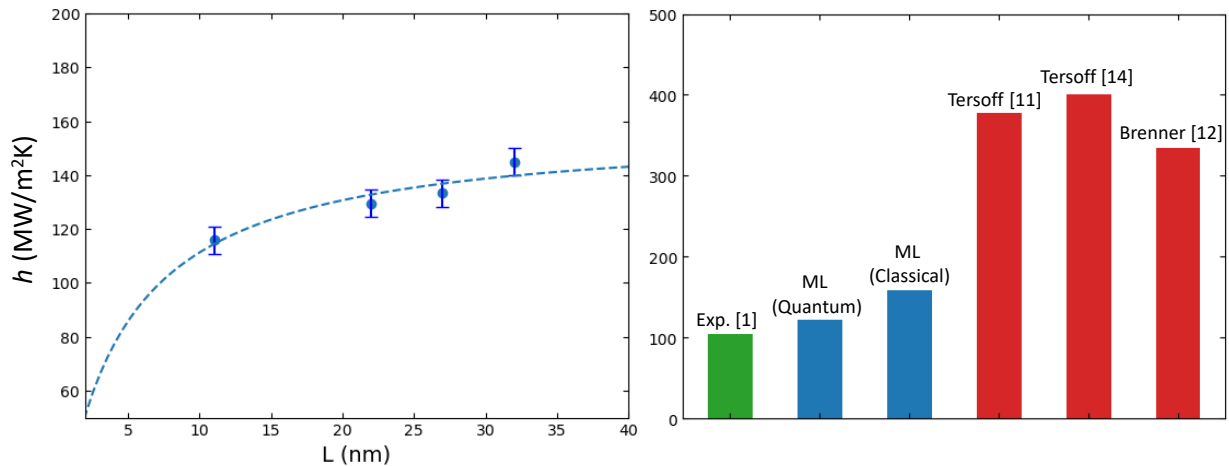

Fig 3. **Left panel**: Interfacial thermal conductance of silicon/diamond heterostructure calculated for different sizes using ML potential with classical approximation. The dashed line shows the fitting curve. **right panel:** Comparison of interfacial thermal conductance measured from experiments [11], calculated from ML (classical and quantum-corrected) and from Tersoff and Brenner interatomic potentials from references [11,12,14].

To interpret the significant difference between the ML and Tersoff potential results, we calculate and plot the spectral interfacial thermal conductance as a function of the phonon frequency. Figure 4(left panel) shows the distribution of the interfacial thermal conductance for the two potentials, ML and Tersoff, using classical and quantum approximation. As can be seen, the Tersoff potential overestimates the value of $h(\omega)$ for all frequencies. Figure 4(right panel) shows the cumulative variation of $h$ with respect to phonon frequency. It shows that for ML potential, the conductance at frequencies below 5 THz and above 17 THz is negligible (comparable to the numerical noise), and also that the highest increase of *h* occurs in the frequency range of 10-15 THz.

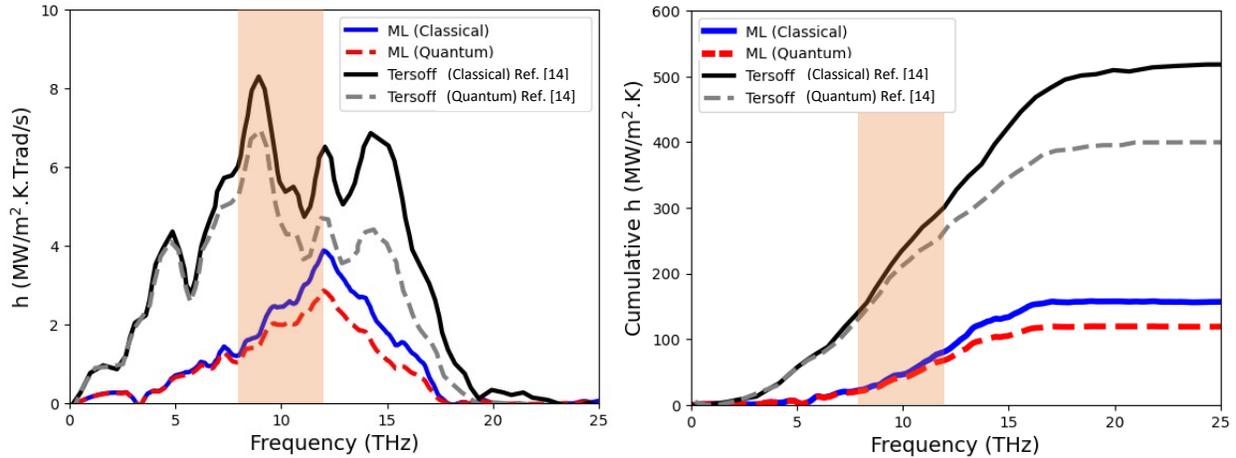

Fig4. **left panel:** Frequency-dependent interfacial thermal conductance of silicon/diamond heterostructure using ML potential and Tersoff potential [14]; **right panel:** Cumulative conductance versus the frequency using ML potential (classical and quantum-corrected). The shaded area are the phonon modes localized at the interface.

Based on the Landauer approach, the interfacial thermal conductance depends on the phonon group velocity and interface bonding, which both affect the phonon transmission at the interface [7,45]. Phonon transmission at interfaces involves also inelastic phonon scattering which depends on the anharmonicities of atomic vibrations in the vicinity of the interface. In the following, we aim at understand why empirical potentials strongly overestimate the interface thermal conductance by separately comparing the accuracy of the group velocities, interfacial bonding and the anharmonicity as predicted by the ML and the empirical potentials retained to model silicon, diamond and the interface. Group velocities will be obtained through bulk phonon dispersion curves. Interfacial bonding will be estimated based on the temperature dependence of the mean square displacements (MSD) of interface atoms. To quantify the amount of vibrational anharmonicities, we will separate anharmonicities in the bulk of the materials that can be evaluated based on phonon lifetimes calculations. The level of anharmonicity of atomic vibrations at the interface may also be measured by the temperature dependence of the MSD of interfacial atoms.

First, we concentrate on the predictions of the group velocities of the bulk materials. Figure 5 shows the phonon dispersion relation of bulk silicon and diamond based on experimental data, DFT calculations, ML potential, and Tersoff potential computations. The Phonopy package was used to calculate the phonon dispersion relation using classical interatomic potentials. As can be seen in Figure 5, the ML potential predicts the phonon dispersion relation in better agreement with the DFT and experimental results. In addition, the Tersoff potential displays high frequencies that are not present in the experimental data and DFT results. By calculating the group velocity of the phonons from the dispersion relation shown in Figure 6, we find that the Tersoff potential overpredicts the phonon group velocity compared to the ML potential with respect to the DFT calculations. Moreover, Figure 6 shows the phonon lifetimes for two bulk silicon and diamond structures. As can be seen, the agreement between the phonon lifetimes calculated from the DFT and ML potential is much better than for the Tersoff potential, which significantly overestimates it, particularly for silicon.

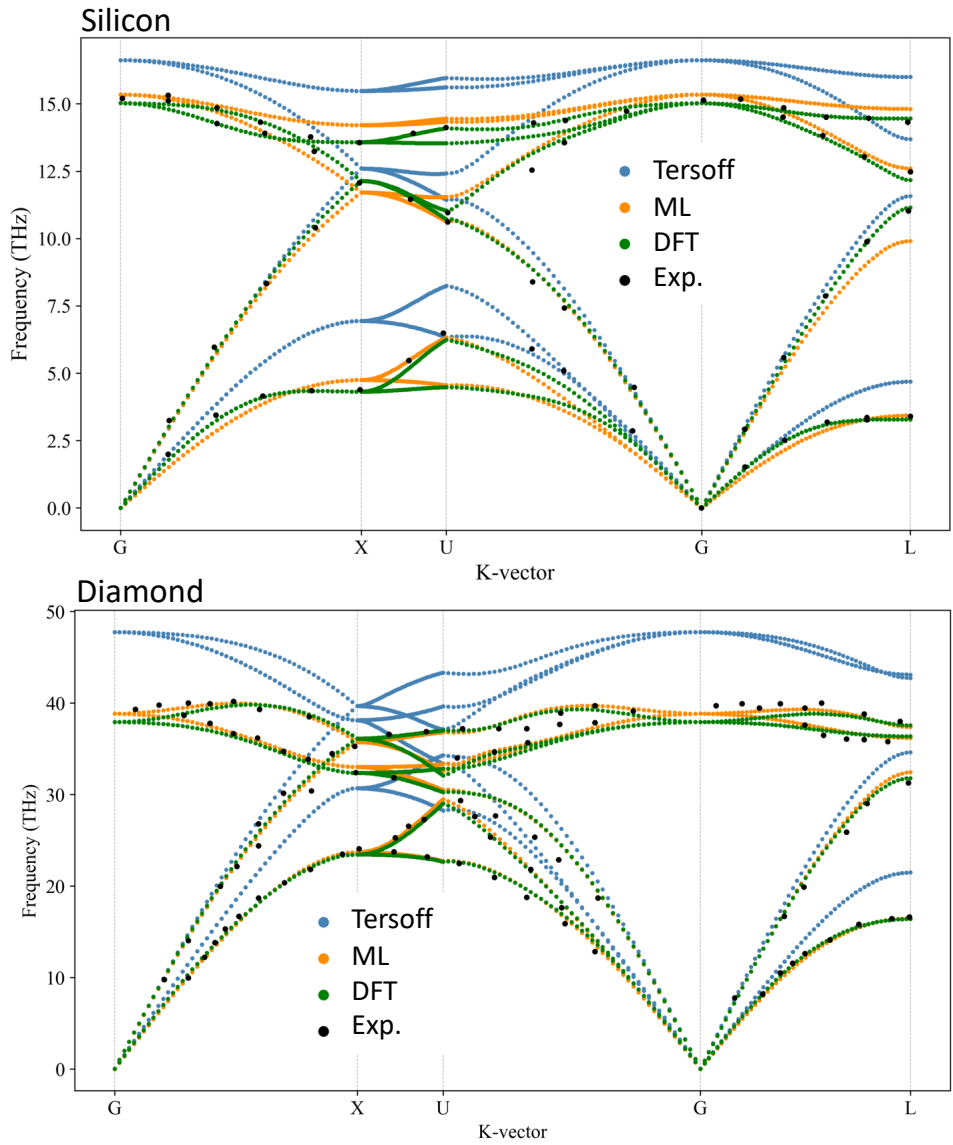

Fig5. Phonon dispersion relation calculated with ML potential and Tersoff potential and compared with DFT

calculations and experimental data [46–49] for bulk silicon (**top panel**) and bulk diamond (**bottom panel**)

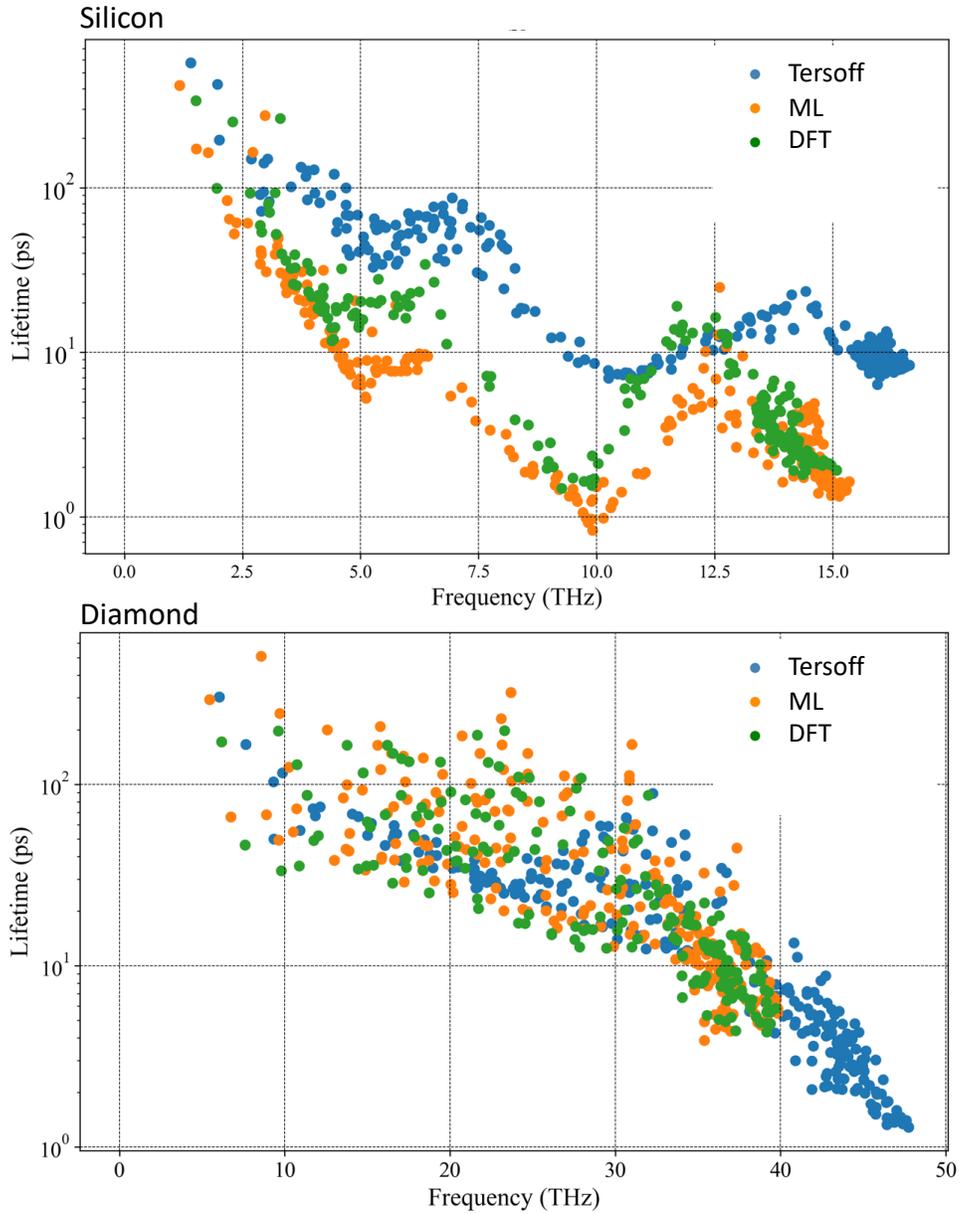

Fig6. Phonon lifetime calculated with ML potential and Tersoff potential and compared with DFT calculations for bulk silicon (**top panel**) and bulk diamond (**bottom panel**)

So far, we have considered the behavior of the bulk systems to interpret why Tersoff potential overestimates the interfacial thermal conductance compared to ML potential. We now investigate the interface properties. To estimate the interfacial bond strength, a region with a thickness of one atomic layer of C on the right side and a region with a thickness of one atomic layer of Si on the left side of the interface are considered. Then the mean square displacement (MSD) of these two regions is calculated using two potentials, ML and Tersoff. Figure 7 shows the results of the MSD versus temperature from 100 K to 400 K. As can be seen, the Tersoff potential predicts a lower value of the MSD of carbon atoms compared to the ML potential. This means that the Tersoff potential corresponds to a stronger bonding at the silicon/diamond interface than the ML potential. This finding is consistent with the higher interfacial thermal conductance of the Tersoff potential compared to the ML potential. Moreover, the nonlinear behavior of MSD of Si atoms versus the temperature variations above 300 K highlights the importance of anharmonic vibrations in thermal transport as a new channel. The inelastic interactions due to anharmonicity in the system lead to higher phonon lifetimes in bulk silicon, as shown for the Tersoff potential in Figure 6.

Additionally, the contribution of anharmonic vibrations to interfacial thermal transport may explain why the Tersoff potential predicts higher interfacial thermal conductance (ITC).

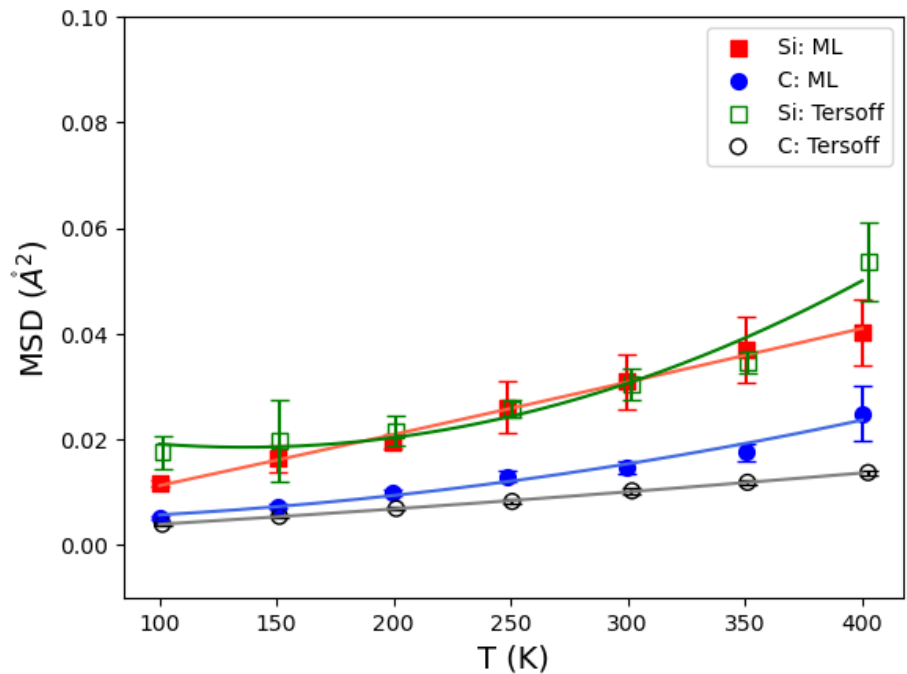

Fig7. Mean-squared displacement of two atomic layers of Si and C atoms adjacent to the interface calculated by Tersoff and ML potentials at various temperatures.

To see which frequencies contribute to heat transport at silicon/diamond interface, we calculate the vibrational density of states (VDOS) in the four different regions of the heterostructure. First, we consider two regions far from the interface with a thickness of 0.5 nm, and we also consider two regions near both sides of the interface having a thickness equal to the ML potential cutoff (0.5 nm). Figure 8 shows the VDOS for these four regions. As can be seen, new frequencies in the range of 8-12 THz, corresponding to localized phonon modes, have appeared at the interface, which are not present in the bulk regions. Based on the accumulative ITC shown in Figure 4, these phonon modes give an important contribution to the interfacial thermal conductance of the silicon/diamond heterostructure in which about 40% of total ITC belongs to this frequency range. Note that the importance of interface modes has been confirmed by simulations and experiments in previous studies for similar silicon/germanium heterostructures [50–52].

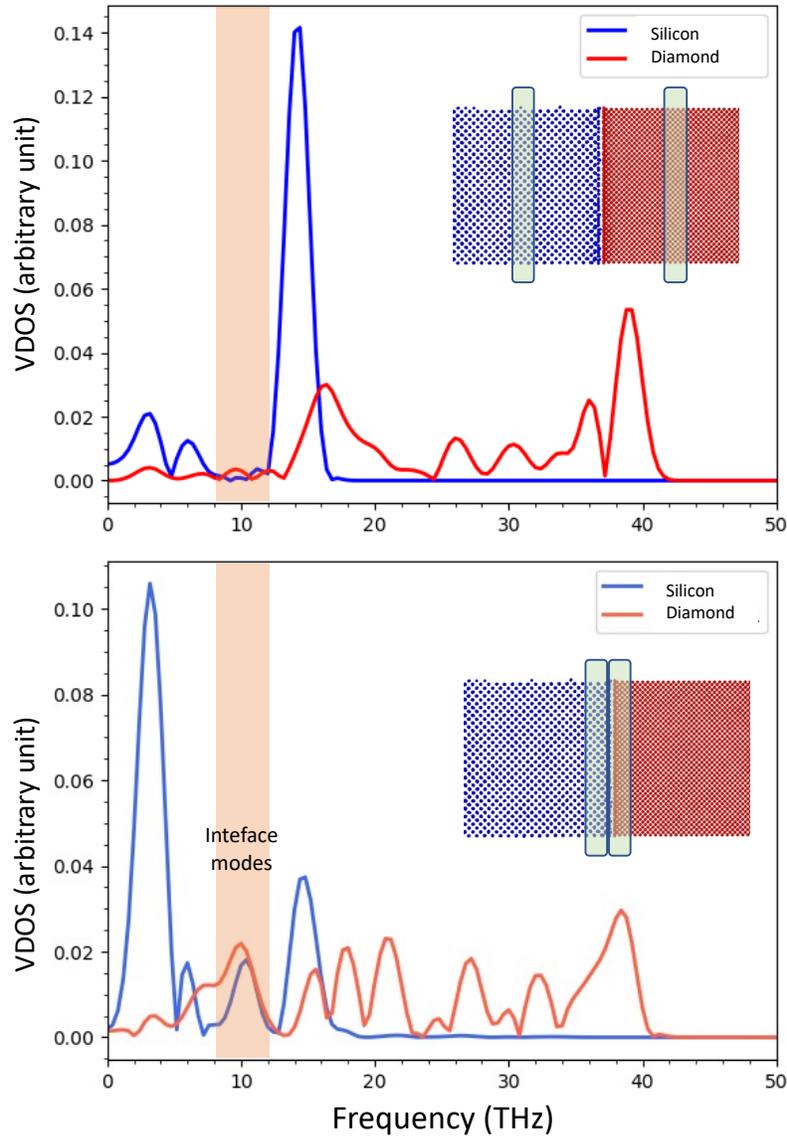

Fig8. Phonon density of states calculated with ML potential for two regions far from the interface (**upper panel**) and for two regions near the interface (**lower panel**). The shaded areas correspond to interface phonon modes that are localized at the interface

**Conclusion**

This study demonstrated the performance of machine learning (ML) interatomic potentials to accurately estimate the interfacial thermal conductance (ITC) between silicon and diamond. By using non-equilibrium molecular dynamics (NEMD) simulations trained on density functional theory (DFT) data, we have addressed the limitations of traditional semi-empirical potentials, such as Tersoff and Brenner, which tend to overpredict ITC by as much as about 300%. Our results show that the ML potentials provide a significantly closer match to experimental ITC values, with a deviation of only 10%, much less than discrepancies observed with semi-empirical potentials. This accuracy is achieved while maintaining computational efficiency, demonstrating the practical applicability of ML potentials in simulating complex thermal transport phenomena at interfaces. The frequency-dependent analysis of heat transfer revealed that phonon interactions play a crucial role in interfacial thermal conductance, and the ML potential effectively captures these interactions. The phonon dispersion relations and lifetimes predicted by the ML potential align well with both DFT calculations and experimental observations, underscoring the potential of ML methods in enhancing our understanding of thermal transport at interfaces. By contrast, we showed that not only empirical potentials do not predict accurately bulk phonon dispersion curves, but they can also do a poor job in describing both interfacial bonding and vibrational anharmonicities of interfacial atoms, which affects the predictions of interface thermal conductance through elastic and inelastic phonon scattering respectively.

Overall, this study highlights the advantages of ML interatomic potentials in providing detailed insights into the atomic-level mechanisms governing thermal transport, which are critical for designing and optimizing high-performance electronic and optoelectronic devices. Future work could extend this approach to other material systems and explore the impact of different interface conditions and configurations, further broadening the applicability and robustness of ML-driven thermal management solutions.

**Acknowledgment**

This research received support from the French Agence Nationale de la Recherche (ANR) under Grant No. ANR-20-CE05-0021-01 (NearHeat). We also acknowledge the use of computational resources from Raptor at iLM, Université Claude Bernard Lyon 1. Also, B. M. and X. Z. appreciate the funding by the Deutsche Forschungsgemeinschaft (DFG, German Research Foundation) under Germany's Excellence Strategy within the Cluster of Excellence PhoenixD (EXC 2122, Project ID 390833453).

**Conflicts of interest**

There are no conflicts of interest to declare.

**Data availability**

The data that support the findings of this study are available upon reasonable request. Please contact the corresponding author for data inquiries.

**References**

[1]    X. Zhu, T. Bi, X. Yuan, Y. Chang, R. Zhang, Y. Fu, J. Tu, Y. Huang, J. Liu, C. Li, H. Kawarada, C-Si interface on SiO2/(1 1 1) diamond p-MOSFETs with high mobility and excellent normally-off operation, Appl Surf Sci 593 (2022) 153368. https://doi.org/10.1016/j.apsusc.2022.153368.

[2]    O.M. Küttel, E. Schaller, J. Osterwalder, L. Schlapbach, X-ray photoelectron diffraction on the nickel/diamond, the silicon/diamond and the gold/diamond interface, Diam Relat Mater 4 (1995) 612–616. https://doi.org/10.1016/0925-9635(94)05216-6.

[3]    Z.B. Milne, J.D. Schall, T.D.B. Jacobs, J.A. Harrison, R.W. Carpick, Covalent Bonding and Atomic-Level Plasticity Increase Adhesion in Silicon–Diamond Nanocontacts, ACS Appl Mater Interfaces 11 (2019) 40734–40748. https://doi.org/10.1021/acsami.9b08695.

[4]    M. Citroni, S. Lagomarsino, G. Parrini, M. Santoro, S. Sciortino, M. Vannoni, G. Ferrari, A. Fossati, F. Gorelli, G. Molesini, G. Piani, A. Scorzoni, A novel method of preparation of silicon-on-diamond materials, Diam Relat Mater 19 (2010) 950–955. https://doi.org/10.1016/j.diamond.2010.02.038.

[5]    G.N. Yushin, A. Aleksov, S.D. Wolter, F. Okuzumi, J.T. Prater, Z. Sitar, Wafer bonding of highly oriented diamond to silicon, Diam Relat Mater 13 (2004) 1816–1821. https://doi.org/10.1016/j.diamond.2004.04.007.

[6]    S. Volz, ed., Thermal Nanosystems and Nanomaterials, Springer Berlin Heidelberg, Berlin, Heidelberg, 2009. https://doi.org/10.1007/978-3-642-04258-4.

[7]    J. Chen, X. Xu, J. Zhou, B. Li, Interfacial thermal resistance: Past, present, and future, Rev Mod Phys 94 (2022) 025002. https://doi.org/10.1103/RevModPhys.94.025002.

[8]  G. Chen, Introduction, in: Nanoscale Energy Transport And Conversion, Oxford University PressNew York, NY, 2005: pp. 3–42. https://doi.org/10.1093/oso/9780195159424.003.0001.

[9]  S. Bin Mansoor, B.S. Yilbas, Phonon transport in silicon–silicon and silicon–diamond thin films: Consideration of thermal boundary resistance at interface, Physica B Condens Matter 406 (2011) 2186–2195. https://doi.org/10.1016/j.physb.2011.03.028.

[10]  K.E. Goodson, O.W. Käding, M. Rösler, R. Zachai, Experimental investigation of thermal conduction normal to diamond-silicon boundaries, J Appl Phys 77 (1995) 1385–1392. https://doi.org/10.1063/1.358950.

[11]  Z. Cheng, T. Bai, J. Shi, T. Feng, Y. Wang, M. Mecklenburg, C. Li, K.D. Hobart, T.I. Feygelson, M.J. Tadjer, B.B. Pate, B.M. Foley, L. Yates, S.T. Pantelides, B.A. Cola, M. Goorsky, S. Graham, Tunable Thermal Energy Transport across Diamond Membranes and Diamond–Si Interfaces by Nanoscale Graphoepitaxy, ACS Appl Mater Interfaces 11 (2019) 18517–18527. https://doi.org/10.1021/acsami.9b02234.

[12]  N. Khosravian, M.K. Samani, G.C. Loh, G.C.K. Chen, D. Baillargeat, B.K. Tay, Molecular dynamic simulation of diamond/silicon interfacial thermal conductance, J Appl Phys 113 (2013). https://doi.org/10.1063/1.4775399.

[13]  L. Chen, S. Chen, Y. Hou, Understanding the thermal conductivity of Diamond/Copper composites by first-principles calculations, Carbon N Y 148 (2019) 249–257. https://doi.org/https://doi.org/10.1016/j.carbon.2019.03.051.

[14]  Y. Qu, J. Yuan, N. Deng, W. Hu, S. Wu, H. Wang, Effect of the thickness of amorphous silicon intermediate layer on the thermal transport of silicon/diamond interface, Results Phys 52 (2023) 106827. https://doi.org/10.1016/j.rinp.2023.106827.

[15]  A. Rajabpour, S. Volz, Universal interfacial thermal resistance at high frequencies, Phys Rev B Condens Matter Mater Phys (2014) 2–5. https://doi.org/10.1103/PhysRevB.90.195444.

[16]  M.S. Alborzi, A. Rajabpour, Effect of overlapping junctions on the heat transfer between 2D layered composite materials, International Communications in Heat and Mass Transfer 109 (2019). https://doi.org/10.1016/j.icheatmasstransfer.2019.104348.

[17]    A. Rajabpour, S. Volz, Thermal boundary resistance from mode energy relaxation times: Case study of argon-like crystals by molecular dynamics, J Appl Phys 108 (2010) 094324. https://doi.org/10.1063/1.3500526.

[18]    S.M. Hatam-Lee, K. Gordiz, A. Rajabpour, Lattice-dynamics-based descriptors for interfacial heat transfer across two-dimensional carbon-based nanostructures, J Appl Phys 130 (2021) 135106. https://doi.org/10.1063/5.0055708.

[19]    Y. Guo, C. Adessi, M. Cobian, S. Merabia, Atomistic simulation of phonon heat transport across metallic vacuum nanogaps, Phys Rev B 106 (2022) 085403–085410. https://doi.org/10.1103/PhysRevB.106.085403.

[20]    H. Han, S. Mérabia, F. Müller-Plathe, Thermal Transport at Solid-Liquid Interfaces: High Pressure Facilitates Heat Flow through Nonlocal Liquid Structuring, Journal of Physical Chemistry Letters 8 (2017) 1946–1951. https://doi.org/10.1021/acs.jpclett.7b00227.

[21]    Z. Fan, Y. Wang, P. Ying, K. Song, J. Wang, Y. Wang, Z. Zeng, K. Xu, E. Lindgren, J.M. Rahm, A.J. Gabourie, J. Liu, H. Dong, J. Wu, Y. Chen, Z. Zhong, J. Sun, P. Erhart, Y. Su, T. Ala-Nissila, GPUMD: A package for constructing accurate machine-learned potentials and performing highly efficient atomistic simulations, J Chem Phys 157 (2022). https://doi.org/10.1063/5.0106617.

[22]    X. Wu, W. Zhou, H. Dong, P. Ying, Y. Wang, B. Song, Z. Fan, S. Xiong, Correcting force error-induced underestimation of lattice thermal conductivity in machine learning molecular dynamics, J Chem Phys 161 (2024). https://doi.org/10.1063/5.0213811.

[23]    K. Xu, Y. Hao, T. Liang, P. Ying, J. Xu, J. Wu, Z. Fan, Accurate prediction of heat conductivity of water by a neuroevolution potential, J Chem Phys 158 (2023). https://doi.org/10.1063/5.0147039.

[24]    Z. Fan, Z. Zeng, C. Zhang, Y. Wang, K. Song, H. Dong, Y. Chen, T. Ala-Nissila, Neuroevolution machine learning potentials: Combining high accuracy and low cost in atomistic simulations and application to heat transport, Phys Rev B 104 (2021) 104309.

[25]    H. Dong, Y. Shi, P. Ying, K. Xu, T. Liang, Y. Wang, Z. Zeng, X. Wu, W. Zhou, S. Xiong, S. Chen, Z. Fan, Molecular dynamics simulations of heat transport using machine-learned

potentials: A mini-review and tutorial on GPUMD with neuroevolution potentials, J Appl Phys 135 (2024). https://doi.org/10.1063/5.0200833.

[26]  A.T. Vu, S. Gulati, P.-A. Vogel, T. Grunwald, T. Bergs, Machine learning-based predictive modeling of contact heat transfer, Int J Heat Mass Transf 174 (2021) 121300. https://doi.org/https://doi.org/10.1016/j.ijheatmasstransfer.2021.121300.

[27]  R. Hu, S. Iwamoto, L. Feng, S. Ju, S. Hu, M. Ohnishi, N. Nagai, K. Hirakawa, J. Shiomi, Machine-learning-optimized aperiodic superlattice minimizes coherent phonon heat conduction, Phys Rev X 10 (2020) 21050.

[28]  S. Arabha, Z.S. Aghbolagh, K. Ghorbani, S.M. Hatam-Lee, A. Rajabpour, Recent advances in lattice thermal conductivity calculation using machine-learning interatomic potentials, J Appl Phys 130 (2021) 210903. https://doi.org/10.1063/5.0069443.

[29]  B. Mortazavi, A. Rajabpour, X. Zhuang, T. Rabczuk, A. V Shapeev, Exploring thermal expansion of carbon-based nanosheets by machine-learning interatomic potentials, Carbon N Y (2022). https://doi.org/https://doi.org/10.1016/j.carbon.2021.10.059.

[30]  B. Mortazavi, I.S. Novikov, A. V Shapeev, A machine-learning-based investigation on the mechanical/failure response and thermal conductivity of semiconducting BC2N monolayers, Carbon N Y 188 (2022) 431–441. https://doi.org/https://doi.org/10.1016/j.carbon.2021.12.039.

[31]  X. Qian, S. Peng, X. Li, Y. Wei, R. Yang, Thermal conductivity modeling using machine learning potentials: application to crystalline and amorphous silicon, Materials Today Physics 10 (2019) 100140. https://doi.org/https://doi.org/10.1016/j.mtphys.2019.100140.

[32]  G. Kresse, J. Furthmüller, Efficient iterative schemes for ab initio total-energy calculations using a plane-wave basis set, Phys Rev B 54 (1996) 11169–11186. https://doi.org/10.1103/PhysRevB.54.11169.

[33]  J.P. Perdew, A. Ruzsinszky, G.I. Csonka, O.A. Vydrov, G.E. Scuseria, L.A. Constantin, X. Zhou, K. Burke, Restoring the Density-Gradient Expansion for Exchange in Solids and Surfaces, Phys Rev Lett 100 (2008) 136406. https://doi.org/10.1103/PhysRevLett.100.136406.

[34] D.J. Chadi, M.L. Cohen, Special points in the brillouin zone, Phys Rev B 8 (1973) 5747–5753. https://doi.org/10.1103/PhysRevB.8.5747.

[35] A.S. Ivan Novikov, Konstantin Gubaev, Evgeny Podryabinkin, The MLIP package: Moment Tensor Potentials with MPI and Active Learning, Mach Learn Sci Technol 2 (2021) 025002.

[36] E. V Podryabinkin, A. V Shapeev, Active learning of linearly parametrized interatomic potentials, Comput Mater Sci 140 (2017) 171–180. https://doi.org/10.1016/j.commatsci.2017.08.031.

[37] B. Mortazavi, E. V Podryabinkin, S. Roche, T. Rabczuk, X. Zhuang, A. V Shapeev, Machine-learning interatomic potentials enable first-principles multiscale modeling of lattice thermal conductivity in graphene/borophene heterostructures, Mater Horiz 7 (2020) 2359–2367. https://doi.org/10.1039/D0MH00787K.

[38] B. Mortazavi, X. Zhuang, T. Rabczuk, A. V Shapeev, Atomistic modeling of the mechanical properties: the rise of machine learning interatomic potentials, Mater Horiz 10 (2023) 1956–1968. https://doi.org/10.1039/D3MH00125C.

[39] A.P. Thompson, H.M. Aktulga, R. Berger, D.S. Bolintineanu, W.M. Brown, P.S. Crozier, P.J. in 't Veld, A. Kohlmeyer, S.G. Moore, T.D. Nguyen, R. Shan, M.J. Stevens, J. Tranchida, C. Trott, S.J. Plimpton, LAMMPS - a flexible simulation tool for particle-based materials modeling at the atomic, meso, and continuum scales, Comput Phys Commun 271 (2022) 108171. https://doi.org/10.1016/j.cpc.2021.108171.

[40] I.S. Novikov, K. Gubaev, E. V Podryabinkin, A. V Shapeev, The MLIP package: moment tensor potentials with MPI and active learning, Mach Learn Sci Technol 2 (2021) 25002. https://doi.org/10.1088/2632-2153/abc9fe.

[41] K. Sääskilahti, J. Oksanen, J. Tulkki, S. Volz, Spectral mapping of heat transfer mechanisms at liquid-solid interfaces, Phys Rev E 93 (2016) 052141. https://doi.org/10.1103/PhysRevE.93.052141.

[42] K. Sääskilahti, J. Oksanen, S. Volz, J. Tulkki, Frequency-dependent phonon mean free path in carbon nanotubes from nonequilibrium molecular dynamics, Phys Rev B 91 (2015) 115426. https://doi.org/10.1103/PhysRevB.91.115426.

[43]    K. Sääskilahti, J. Oksanen, J. Tulkki, S. Volz, Role of anharmonic phonon scattering in the spectrally decomposed thermal conductance at planar interfaces, Phys Rev B 90 (2014) 134312. https://doi.org/10.1103/PhysRevB.90.134312.

[44]    Y. Guo, M. Gómez Viloria, R. Messina, P. Ben-Abdallah, S. Merabia, Atomistic modeling of extreme near-field heat transport across nanogaps between two polar dielectric materials, Phys Rev B 108 (2023) 085434. https://doi.org/10.1103/PhysRevB.108.085434.

[45]    E.T. Swartz, Thermal boundary resistance, Rev. Mod. Phys 61 (n.d.) 605–668,.

[46]    J.L. Warren, J.L. Yarnell, G. Dolling, R.A. Cowley, Lattice Dynamics of Diamond, Physical Review 158 (1967) 805–808. https://doi.org/10.1103/PhysRev.158.805.

[47]    M. Schwoerer-Böhning, A.T. Macrander, D.A. Arms, Phonon Dispersion of Diamond Measured by Inelastic X-Ray Scattering, Phys Rev Lett 80 (1998) 5572–5575. https://doi.org/10.1103/PhysRevLett.80.5572.

[48]    J.Q. Broughton, X.P. Li, Phase diagram of silicon by molecular dynamics, Phys Rev B 35 (1987) 9120–9127. https://doi.org/10.1103/PhysRevB.35.9120.

[49]    X.-P. Li, G. Chen, P.B. Allen, J.Q. Broughton, Energy and vibrational spectrum of the Si(111) (7×7) surface from empirical potentials, Phys Rev B 38 (1988) 3331–3341. https://doi.org/10.1103/PhysRevB.38.3331.

[50]    Z. Cheng, R. Li, X. Yan, G. Jernigan, J. Shi, M.E. Liao, N.J. Hines, C.A. Gadre, J.C. Idrobo, E. Lee, K.D. Hobart, M.S. Goorsky, X. Pan, T. Luo, S. Graham, Experimental observation of localized interfacial phonon modes, Nat Commun 12 (2021) 6901. https://doi.org/10.1038/s41467-021-27250-3.

[51]    R.K. Kelayeh, A. Rajabpour, E. Taheran, Y. Bahari, Optimization of interfacial mixing for thermal transport along Si/Ge heterostructures: A molecular dynamics study, Appl Surf Sci 626 (2023) 157236. https://doi.org/10.1016/j.apsusc.2023.157236.

[52]    Y. Guo, Z. Zhang, M. Bescond, S. Xiong, M. Nomura, S. Volz, Anharmonic phonon-phonon scattering at the interface between two solids by nonequilibrium Green's function formalism, Phys Rev B 103 (2021) 174306. https://doi.org/10.1103/PhysRevB.103.174306.